\documentclass[11pt]{article}
\usepackage{fancyhdr}
\usepackage{isomath}
\usepackage{amsmath}
\usepackage{amsbsy}
\usepackage{amssymb}
\usepackage{amscd}
\usepackage{amsfonts}
\usepackage{graphicx,color}
\usepackage{verbatim}
\usepackage{euscript}
\usepackage{alltt}
\usepackage{stmaryrd}
\usepackage{subfigure}
\usepackage{relsize}
\usepackage{enumitem}
\usepackage{xcolor}

\DeclareGraphicsExtensions{.eps,.pdf}

\usepackage[margin=1in]{geometry}

\newcommand{\veps}{\varepsilon}
\newcommand{\p}{\partial}

\date{}
\begin{document}
\title{An action for nonlinear dislocation dynamics}

\author{Amit Acharya\thanks{Department of Civil \& Environmental Engineering, and Center for Nonlinear Analysis, Carnegie Mellon University, Pittsburgh, PA 15213, email: acharyaamit@cmu.edu.}}
\maketitle

\begin{abstract}
\noindent An action functional is developed for nonlinear dislocation dynamics. This serves as a first step towards the application of effective field theory in physics to evaluate its potential in obtaining a macroscopic description of dislocation dynamics describing the plasticity of crystalline solids. Connections arise between the continuum mechanics and material science of defects in solids, effective field theory techniques in physics, and fracton tensor gauge theories. 

The scheme that emerges from this work for generating a variational principle for a nonlinear pde system is  general, as is demonstrated by doing so for nonlinear elastostatics involving a stress response function that is not necessarily hyperelastic.
\noindent 
\end{abstract}

\section{Introduction}\label{sec:intro}
The goal of this work is to develop a setting for enabling the application of methods of Effective Field theory (EFT), as described in \cite{zaanen2004duality, beekman2017dual1, beekman2017dual, kleinert1989gauge_1, kleinert1989gauge_2}, to the study of nonlinear dislocation dynamics, posed as a system of nonlinear partial differential equations (pde) as in \cite{acharya2004constitutive, acharya2011microcanonical}. The adopted strategy is to explore action functionals that correspond to the given system of pde in some sense to be made precise in each case.
A variational perspective often allows systematic ways of approaching approximations to a problem (through bounds - when one has minimum/maximum principles - and by relaxing regularity requirements on the solution to the problem, e.g.), and the hope is also that, assuming that the action-based state-space-measure typically invoked in path-integral methods is relevant to the physical study of dislocation dynamics, a start on one approach to studying fluctuations and renormalization in the subject can be made. Some idea of what can be expected in terms of fluctuations, and the need for coarse-graining/renormalization in nonlinear dislocation dynamics can be obtained from the results presented in \cite{arora2020unification, arora2020finite, arora_acharya_ijss}. From the point of view of continuum mechanics and materials science, it suffices to study (approximate) solutions to the pde system itself, the nonlinear dynamics of which may display complex behavior in limiting situations with similarities to stochastic response. The Effective Field theory perspective, where statistical properties in the form of successively higher-order (space-time) correlation functions of the fields describing a physical system are studied \cite{kleinert1989gauge_1}, has as a \textit{prerequisite} the definition of the problem in the form of an action functional. Initiating a bridge between these points of view is the main motivation of this work.

It is well understood that not every system of pde admits a variational principle whose Euler-Lagrange (E-L) equation is the system in question. Even when this is possible in principle, finding such a principle for a generally nonlinear system of pde is a non-trivial task. In this work, we are able to approach the goal in two `relaxed' contexts. In the first, an action functional is developed, \textit{some} of whose E-L equations correspond to those of the pde system of nonlinear dislocation dynamics, this by invoking a change of variables with an associated assumption for its validity. The second case also invokes a change of variables, now without a precondition, but it is only possible to show that a solution of its E-L equations defines a solution of the desired pde system by making an allowed, but special, choice of some of the latter's ingredient fields (that may not necessarily be optimal from the physical point of view). These issues are made clear in the context of the development.

Developing a variational principle for nonlinear dislocation mechanics and elasticity in the spatial setting of continuum mechanics is a non-standard enterprise - in this, our work is inspired by the work of Seliger and Whitham \cite{seliger1968variational} who treat the case of nonlinear elasticity but not dislocations. Due to the fundamental incompatibility of the elastic reference with being a coherent reference in ambient Euclidean space in the presence of defects in the body, Seliger and Whitham's ideas do not naturally extend to our case and, in fact, our considerations provide an essentially different variational formulation from that of \cite{seliger1968variational} for nonlinear elasticity. However, a significant clue their work provides is to look for an `elimination' of the velocity field which is exploited in our work, but not by utilizing an E-L equation of a primal variational principle as done in \cite{seliger1968variational}. Instead, our approach connects naturally to the idea of dualizing a variational principle as practiced in EFT (e.g. \cite{garcia2018multipole}), only here we are able to employ a `partial dualization' because of the nonconvexity of the (strain) energy density in the geometrically nonlinear setting; this has the flavor of a `mixed' variational principle, commonly employed in mechanics, optimization theory, and in the theory of finite element numerical approximations of problems that admit a variational formulation. To our knowledge, a variational principle for nonlinear dislocation dynamics formulated in the spatial setting does not currently exist. Lazar \cite{lazar2011fundamentals} has formulated a gauge theory of dislocations based on the reference configuration; as mentioned, a physically distinguished coherent elastic reference configuration for a solid does not exist in the presence of dislocations - nevertheless, what relation might exist between the gauge theory of Lazar and the current work is a topic worthy of examination in its own right.

An outline of the paper is follows: In Sec.~\ref{sec:FDM} we introduce some notation and the basic equations of the theory of dislocation dynamics we work with and its relation to the theory of nonlinear elasticity as a simplification. Sec.~\ref{sec:main_action} lists the proposed action functionals and demonstrates that their Euler-Lagrange equations have the properties mentioned earlier. Sec.~\ref{sec:primal_action} provides motivation and the basis for the actions proposed in Sec.~\ref{sec:main_action}. In Sec.~\ref{sec:fracton} contact is established between the dynamic extension, in $(3 + 1)$-d, of the classical geometrically linear theory of defects due to DeWit \cite{dewit1971relation, dewit1973theory, dewit1973theory4} and Kr\"oner \cite{kroner1981continuum} as reviewed in Appendix \ref{app:app}, and the theory of fractons \cite{pretko2018fracton}, by developing an appropriate action functional. Sec.~\ref{sec:obs} is a discussion of implications of this work and potential directions for future work; Sec.~\ref{sec:dual_elast} demonstrates the application of the developed ideas for generating a variational principle in the context of nonlinear elastostatics for a Cauchy elastic material.
\section{Equations of Field Dislocation Mechanics}\label{sec:FDM}
In what follows, all tensor indices range from 1 to 3 (spatial). Time is treated as separate from the space variables and denoted by $t$. We exclusively utilize only a rectangular Cartesian coordinate system, and all tensor indices are w.r.t.~the orthonormal basis of this system; the letter $t$ is never used as a tensor index. We refer to the inverse elastic distortion field as $W_{ij}$, $v_i$ refers to the material velocity field, and $V_i$ to the dislocation velocity field. $\rho$ is the mass density. A superposed dot represents a material time derivative. Any spatial domain for the body is assumed to be simply-connected. $\Omega$ will be a fixed spatial domain in ambient 3-d Euclidean space, and $[0,T]$ a fixed interval of time. We will use the shorthand $\psi'_{ij} := \p_{W_{ij}} \psi$ and $\psi''_{ijmn} := \p_{W_{ij}} \p_{W_{mn}} \psi$ which is symmetric under interchange of the pairs $(ij), (mn)$. The curl of a tensor field is understood in terms of row-wise curls; the cross-product of a tensor and a vector corresponds to row-wise cross-products (these operations have invariant meanings). The inclusion of a body force density field is straightforward and requires no particular special consideration, and is not included here without loss of essential generality.

 The inverse elastic distortion, $W$ is a `two-point tensor' in the sense that it maps vectors from (tangent spaces in) the current configuration to a fixed vector space, the latter not altered by superimposed rigid body motions of the body. Its negative $\mbox{curl}$, the dislocation density tensor $\alpha$, is also a tensor of the same nature. Their first indices may be considered `material' in this sense and their second index `spatial.'

The physical statements of conservation of mass, linear momentum, and Burgers vector (the topological charge of dislocations) imply the equations of field dislocation dynamics given by
\begin{subequations}\label{eq:FDM}
\begin{align}
\dot{W}_{ij} + W_{ik} \p_j v_k = \p_t W_{ij} + v_k \p_k W_{ij} + W_{ik} \p_j v_k & = e_{jrs} \alpha_{ir} V_s +  \p_j f_i=: (\alpha \times V)_{ij} + \p_j f_i \label{eq:v-W_comp}\\
\dot{\rho} + \rho \p_k v_k = \p_t \rho +  \p_k (\rho v_k)  & = 0 \label{eq:bom}\\
\rho \dot{v}_i = \rho (\p_t v_i + v_k \p_k v_i) = \p_t (\rho v_i) + \p_j (\rho v_i v_j) & = \p_j (- \rho W_{ki} \psi'_{kj}) \label{eq:blm}
\end{align}
\end{subequations}
where the middle equality in \eqref{eq:blm} assumes that \eqref{eq:bom} holds.

The vector field $f_i$ may be assigned freely without interfering with the conservation of topological charge, but does affect the stress and material velocity fields. Thus, it has to be assigned for a well-set problem \cite{acharya2019structure}, and a physically justified assumption - for a microscopic model representing the situation when all dislocation lines are resolved and plastic deformation is related \textit{entirely} to the motion (or its absence), relative to the material, of this population - is to choose it to vanish \cite{acharya2015dislocation, dzyaloshinskii1980poisson}.

The fields $(\alpha, V)$ - the dislocation density and the dislocation velocity, respectively - can be thought of as specified space-time fields as one option, in which case the system \eqref{eq:FDM} corresponds to a set of equation for the determination of the fields $(W,\rho, v)$, forced by the specified fields (and initial and boundary conditions). Even more interesting dynamics results when the fields $\alpha, V$ are defined in terms of functions of (pointwise) values of $W, \mbox{curl}\,W$, and $\mbox{curl}\,\mbox{curl}\,W$. Both possibilities are considered in this work, see Sec.~\ref{sec:main_action} and observation \ref{itm:obs_1} of Sec.~\ref{sec:obs}.

The statement \eqref{eq:v-W_comp} is the statement of compatibility of the rate of change of the inverse elastic distortion and the particle velocity gradient in the presence of permanent strain rate produced by the motion of dislocations \cite{acharya2015dislocation} (in the parlance of plasticity theory, it provides a kinematically fundamental basis for an `additive strain rate decomposition' of the particle velocity gradient into elastic and plastic parts). Some intuition for the term $\alpha \times V$ is as follows: consider the special form  $\alpha := b \, \otimes \,l$ where $b$ represents the Burgers vector of a dislocation curve with tangent direction $l$ and let the curve be moving with the velocity field $V$ w.r.t the material. Then $l \times V$ in $\alpha \times V = b \otimes ( l \times V)$ represents the (space-like part of) an element of the `world-sheet' of the moving dislocation transporting its topological charge $b$; for a mechanical interpretation, this produces a permanent/plastic strain rate in the direction of $b$ across surfaces with normal in the direction $(l \times V)$ - for $b$ belonging to the surface, this is a shear strain rate. Statements \eqref{eq:bom} and \eqref{eq:blm} represent the balances of mass and linear momentum, respectively. The term  $- \rho W_{ki} \psi'_{kj}$ represents the Cauchy stress tensor $T_{ij}$, which can be shown to be symmetric due to invariance under superposed rigid motions of the function $\psi(W)$. Hence, balance of angular momentum is also satisfied. Modeling the scale-invariance (over a wide range of length scales) of purely elastic response, the existence of lattice-invariant (non-trivial) deformations, and the invariance under superposed rigid deformations of the strain energy density function $\psi$ implies that it is necessarily non-convex in $(WW^T)^{-1}$, the latter known in nonlinear elasticity as the elastic Right Cauchy-Green tensor; simply invariance under superposed rigid deformations implies that $\psi$ when viewed as a function of $W$ (through $WW^T$) cannot be convex (as discussed in Sec. \ref{sec:obs}), a fact that prevents invoking a Legendre transform for it\footnote{The practitioner of plasticity theory typically is used to a set of mechanical governing equations consisting of balance of mass, linear momentum (and angular momentum, satisfied by a symmetric stress tensor), and an additive decomposition of the velocity gradient into elastic and plastic parts, with the plastic part specified through a constitutive equation in terms of stress, along with a constitutive equation for the stress itself. 

In the system \eqref{eq:FDM}, \eqref{eq:v-W_comp} is exactly the said decomposition (which can be seen in more familiar notation by using $F^e := W^{-1}$) with its rhs representing the plastic strain rate, which is either a specified function or a function of $W$, $\mbox{curl} W$, and $\mbox{curl}\,\mbox{curl}\,W$ through the stress, the dislocation density and its $\mbox{curl}$, the latter specification in accord with positive dissipation. Such a model is capable of representing localized large deformations produced due to the motion of individual dislocations and phenomena like metastable equilibria of single dislocations with compact cores and their annihilation and dissociation, as demonstrated in \cite{zhang2015single,arora2020finite}. There is an attractive bare-bones nature to this microscopic theory with time scales set by elastic wave propagation, single dislocation mobility encoded in the constitutive equation for $V$ in terms of its theoretically defined thermodynamic driving force, and rate of loading, and an intrinsic material length scale arising from the dependence of the potential $\psi$ on the dislocation density, modeling dislocation core energy. Slip-system like behavior arises from the lattice symmetries - anisotropy and periodicity - encoded in the potential $\psi$, which can be modeled directly from interatomic potentials \cite{milstein1980theoretical,chantasiriwan1996higher}. Nonlinearity of the field equations coupled to these simple ingredients gives rise to complex interactions and dynamics. Understanding the detailed behavior of such a microscopic theory and its upscaling remains a grand-and-glorious unfulfilled goal, to which we want to bring the tools of EFT and evaluate them.}.

We note that the system \eqref{eq:FDM} comprises $(9+1+3)$ equations in the $13$ fields $(W_{ij}, \rho, v_i)$. While not strictly necessary for the purpose of this work, we note in passing the following important characteristics of the system \eqref{eq:FDM}, either demonstrated in, or easily deduced from, \cite{acharya2004constitutive, acharya2011microcanonical}. If the fields $(\alpha, V)$ are constrained to satisfy 
\begin{equation}\label{eq:burgers_conserv_law}
\overset{\circ}{\alpha} = - \mbox{curl}({\alpha} \times V)
\end{equation}
with $\overset{\circ}{\alpha}_{ij} = (\p_k v_k) \alpha_{ij} + \dot{\alpha}_{ij} - \alpha_{ik} \p_k v_j$, then \eqref{eq:v-W_comp} implies $ - \mbox{curl}\, W = \alpha$ at all times provided it is initially satisfied. Conversely, if  $ - \mbox{curl}\, W = \alpha$ at all times, then \eqref{eq:v-W_comp} implies the conservation law  \eqref{eq:burgers_conserv_law} (for Burgers vector). Finally, it follows from the last claim that when only $V = 0$, \eqref{eq:v-W_comp} and the definition $\alpha := - \mbox{curl}\,W$ implies that $\overset{\circ}{\alpha} = 0$; the field $\alpha$ evolves in this case but maintaining the constraint that the Burgers vector content of \textit{any} area patch $A$ of the body consisting of the same set of material particles in time remains constant; this Burgers vector content of the patch $A$ is given by $\int_A \alpha_{ij}\nu_j da$, where $\nu$ is the unit normal field on the evolving `material' patch. This last situation is interpreted as the time-dependent, but \textit{elastic}, theory of continuously distributed dislocations. The equation \eqref{eq:v-W_comp} for $V = 0$ is identical to \cite[6.15]{dzyaloshinskii1980poisson}, in the absence of dissipation, the latter obtained by Poisson bracket techniques. In \cite{dzyaloshinskii1980poisson} no commitment is made about the form of the dissipation due to dislocation motion. In \cite{acharya2004constitutive, acharya2011microcanonical} the form of the dislocation current (w.r.t the material) leading to dissipation is derived based on the fundamental nonlinear kinematics of dislocation motion and continuum thermodynamic arguments, leading to the derivation of the  thermodynamic driving force for $V_i$ as given by 
\[
 f^V_i (T, W, \alpha) := e_{irs} T_{jr} (W^{-1})_{jk} (-e_{slm} \p_l W_{km}) =  T_{jm} (W^{-1})_{jk}( \p_m W_{ki} -\p_i W_{km} )
\]
(the `Peach-Koehler force' on a dislocation). Recalling those arguments briefly, when $V_i$  is assumed to be in the direction of this driving force as a constitutive assumption (e.g. $V_i = m f_i^V, m > 0$ a scalar mobility constant), it is seen that the power supplied by the external tractions on a body minus the rate of change of free energy and kinetic energy can be expressed as
\begin{equation*}
\begin{aligned}
\int_{\p \mathcal{D}_t} da \, (T_{ij}n_j) v_i - \frac{d}{dt} \int_{\mathcal{D}_t} d^3x \, \rho \left( \frac{1}{2}  v_i v_i + \psi(W) \right) & = \int_{\mathcal{D}_t} d^3x \, T_{ij} \p_j v_i - \rho \dot{\psi} \\
  & = \int_{\mathcal{D}_t} d^3x \, \left(  T_{ij} -  \rho\, ([\p_{F^e}\, \psi (C^e)]\, F^{eT})_{ij} \right) \p_j v_i  \\
  & \quad + \int_{\mathcal{D}_t} d^3x \, f^V_i (T, W, \alpha) V_i \geq 0,
\end{aligned}
\end{equation*}
i.e. the \textit{mechanical dissipation}, is non-negative for all processes satisfying the system \eqref{eq:FDM} \cite{acharya2004constitutive,acharya2011microcanonical}, where $\mathcal{D}_t$ is the time-parametrized deforming body along a process  and $n_i$ is the outward unit normal field on the boundary; the model is \textit{dissipative} in this sense. In the expression above for mechanical dissipation, use has been made of the thermodynamic stress relation $T = \rho\, [\p_{F^e}\, \psi (C^e)]\, F^{eT} = \rho\, 2 F^e \,[\p_{C^e}\, \psi (C^e)]\, F^{eT}$ which arises as follows.  Frame-indiference of the free energy density requires $\psi(W) = \psi(WQ^T)$ for all orthogonal tensors $Q$, which implies $\psi$ can only be a function of $W W^T$, which can also be stated equivalently as that $\psi$ is a function of $C^e = F^{eT}F^e = (WW^T)^{-1}, F^e := W^{-1}$. Requiring no dissipation in elastic processes (i.e. processes with no dislocation motion, i.e. $V_i = 0$) then provides the abovementioned stress relation (showing also the mechanical dissipation density, a scalar, is objective and unaffected by the material spin, the antisymmetric part of $\mbox{grad}\, v$).
As concerns frame-indifference considerations for \eqref{eq:v-W_comp}, since it follows from localizing an integral statement of balance on the current configuration 
\[
\mbox{curl} W = - \alpha \Longrightarrow \frac{d}{dt} \int_{\p A} W dx = - \frac{d}{dt} \int_A \alpha \nu da =  \int_{\p A} \alpha \times V dx
\]
which holds for all area patches $A$ with unit normal field $\nu$ in the body (with the flux term in the last line integral physically justified in \cite{acharya2011microcanonical}), it automatically satisfies invariance under superposed rigid body motions (as the output of the line integrals are not affected by a superposed time-dependent rigid motion of the body). Indeed, for a superposed rigid body motion characterized by any time dependent rotation tensor valued function $Q(\cdot)$ in which $W$ transforms to $WQ^T$, it is a straightforward check that the quantity $\dot{W} + W \mbox{grad} v$ transforms similarly to $(\dot{W} + W \mbox{grad} v)Q^T$ as does $\alpha \times V$, where $V$ is a spatial vector field.

\subsection{Reduction to nonlinear elasticity}
Nonlinear elasticity is obtained as a special case of \eqref{eq:FDM} by assuming the field  $\alpha \times V = 0$ in \eqref{eq:v-W_comp}. It can then be shown that provided $e_{jkl} \p_k W_{il} = 0$ holds as an initial condition, the same condition holds for all time. Then, there exists functions $\Phi_i$ that satisfy
\begin{equation}\label{eq:a=0}
W_{is} = \p_s \Phi_i.
\end{equation}
Invoking any arbitrarily chosen fixed-in-time reference configuration for the body with points denoted as $X_i$, the definition $\Phi_i^{(r)} (X, t) := \Phi_i (x(X,t),t)$ (suppressing indices when obvious) and the (standard) assumption that $\p_{X_j} \Phi^{(r)}_i (X,t)$ and $\p_{X_j} x_i (X,t)$ have positive determinants so that they are also invertible,
\[
\begin{aligned}
 \p_{X_{k}} \Phi^{(r)}_i \p_{x_{j}} X_k  = \p_{x_j} \Phi_i  \Longrightarrow \dot{\overline{\p_{x_{j}} \Phi_i}}& = \p_{X_k} \left(\p_t \Phi^{(r)}_i \right) \p_{x_j} X_k + \p_{X_k} \Phi^{(r)}_i \dot{\overline{\p_{x_j} X_k}}\\
 &= \p_{X_k} \left(\p_t \Phi^{(r)}_i \right) \p_{x_j} X_k - \p_{x_k} \Phi^{(r)}_i \p_j v_k.
\end{aligned}
\]
Then, \eqref{eq:v-W_comp} and \eqref{eq:a=0} imply
\[ 
\p_{X_k} \left(\p_t \Phi^{(r)}_i \right) \p_{x_j} X_k = 0 \Longrightarrow \p_t \p_{X_k} \Phi^{(r)}_i = 0
\]
which further implies that $\Phi_i^{(r)}$ is a rigid (possibly time-varying) translation of a deformation of the configuration represented by $X$, a translation that can be ignored in the context of elasticity without loss of generality. This means that when $\alpha = 0$, \eqref{eq:FDM} implies the existence of a fixed global stress-free configuration from which the elastic distortion $W^{-1}$ is measured, and is a genuine deformation gradient (with $W$ being the gradient of the inverse deformation). This, along with \eqref{eq:blm} and \eqref{eq:bom} describes nonlinear elasticity theory.
\section{Actions corresponding to nonlinear dislocation dynamics}\label{sec:main_action}
In this section, two actions are developed that correspond to the system \eqref{eq:FDM} in different senses, as alluded to in Sec.~\ref{sec:intro}.
Consider first the action
\begin{equation}\label{eq:mixed_action_1}
    \begin{aligned}
    S_1[A, W, \rho, \theta, \lambda; \alpha, V] = \int_{\Omega \times [0,T]} dt d^3x & -\frac{1}{2} p_i \mathbb{D}_{ij}  p_j - \rho \psi(W) \\
    & + A_{ij}[ \p_t W_{ij} - (\alpha \times V)_{ij} ]\\
    & + \theta \p_t \rho + \lambda_i [ \p_j (-\rho W_{ki} \psi'_{kj} )],
    \end{aligned}
\end{equation}
where 
\begin{equation}\label{eq:def_p_1}
\begin{aligned}
p_k & := -[ A_{ij} \p_k W_{ij} - \p_j (A_{ij} W_{ik}) - \rho \p_k \theta + \rho \p_t \lambda_k ]\\
\mathbb{L}_{ij} & := \rho (\delta_{ij} + \p_j \lambda_i + \p_i \lambda_j) = \mathbb{L}_{ji}\\
\mathbb{D}_{ij} & := \mathbb{L}^{-1}_{ij} = \mathbb{D}_{ji}
\end{aligned}
\end{equation}
and the fields $\alpha_{ij}$ and $V_i$ are considered known functions of $(x,t)$; we consider interesting generalizations in Sec.~\ref{sec:obs}. We have in mind here the situation when $f_i = 0$ in \eqref{eq:FDM}.

\textit{An important assumption here is that $\mathbb{L}$ is an invertible matrix for all possible values of its argument fields}. This is not entirely satisfactory; it can be checked that $\lambda_i$ has to have physical dimensions of length so that the terms beyond the identity tensor in the definition of $\mathbb{L}$ are dimensionless, and most easily associated with a strain, and the assumption is valid for small values of this added `strain.' Also, if the convective nonlinearity in the particle velocity field is ignored then the considerations of Sec.~\ref{sec:primal_action} show that $\mathbb{L}$ could be defined without the added terms; thus it may be reasonable to expect $\mathbb{L}$ to be invertible in motions where the convective nonlinearity in the particle velocity is `small.' 

The first variation of $S_1$, assuming all variations vanish on the boundary of $\Omega \times [0,T]$\footnote{Here, we are interested in interior field equations; natural boundary conditions can be inferred in standard fashion by not assuming the variations to vanish on the boundary.} is given by
\begin{equation}\label{eq:first_var_1_1}
\begin{aligned}
\delta S_1 = \int_{\Omega \times [0,T]} dt d^3x \  & - \frac{1}{2} p_i p_j \delta \mathbb{D}_{ij} - \psi \delta \rho - \rho \psi'_{ij} \delta W_{ij}  \\
& + \delta A_{ij} [ \p_t W_{ij} - (\alpha \times V)_{ij}] - \p_t A_{ij} \delta W_{ij} \\
& + \delta \theta \p_t \rho - \p_t \theta \delta \rho + \delta \lambda_i  \p_j (- \rho W_{ki} \psi'_{kj}) \\
& + \p_j \lambda_i \left[ \delta \rho W_{ki} \psi'_{kj} + \rho \delta W_{ki} \psi'_{kj} + \rho W_{ki} \psi''_{kjmn} \delta W_{mn}\right]\\
& \qquad \qquad \qquad + \\
& \ {\color{blue} - \ p_i \mathbb{D}_{ik} \delta p_k  },
\end{aligned}
\end{equation}
where
\begin{equation}\label{eq:del_D}
\delta \mathbb{D}_{ij} = - \mathbb{D}_{ia} \delta \mathbb{L}_{ab} \mathbb{D}_{bj} \Longrightarrow \delta \mathbb{D}_{ij} = - \frac{\delta \rho}{\rho} \mathbb{D}_{ij} - \mathbb{D}_{ia} \mathbb{D}_{bj} \rho ( \p_b \delta \lambda_a +  \p_a \delta \lambda_b )
\end{equation}
and the contribution to the first variation from the term in blue in \eqref{eq:first_var_1_1} is
\begin{equation}\label{eq:first_var_2_2}
\begin{aligned}
\int_{\Omega \times [0,T]} dt d^3x & \ p_r \mathbb{D}_{rk} \p_k W_{ij} \delta A_{ij} - \p_k  (p_r \mathbb{D}_{rk} A_{ij}) \delta W_{ij} + \p_j (p_r \mathbb{D}_{rk} ) W_{ik} \delta A_{ij} + \p_j (p_r \mathbb{D}_{rk}) A_{ij} \delta W_{ik}\\
& - p_r \mathbb{D}_{rk} \p_k \theta \delta \rho + \p_k( \rho p_r \mathbb{D}_{rk} ) \delta \theta + p_r \mathbb{D}_{rk} \p_t \lambda_k \delta \rho - \p_t ( \rho p_r \mathbb{D}_{rk}) \delta \lambda_k.
\end{aligned}
\end{equation}
Thus, \eqref{eq:first_var_1_1}-\eqref{eq:del_D}-\eqref{eq:first_var_2_2}, using the definitions \eqref{eq:def_p_1}, imply the Euler-Lagrange equations,
\begin{equation}\label{eq:E-L_1}
\begin{aligned}
\delta A_{ij}: \quad & \p_t W_{ij}  +  p_r \mathbb{D}_{rk} \p_k W_{ij} + W_{ik} \p_j(p_r \mathbb{D}_{rk} ) - (\alpha \times V)_{ij} = 0\\
\delta \theta: \quad & \p_t \rho + \p_k ( \rho\, p_r \mathbb{D}_{rk} ) = 0\\
\delta \lambda_k : \quad &  - \p_t (\rho \, p_r \mathbb{D}_{rk} ) - \frac{1}{2} \p_j ( \rho \, p_i \mathbb{D}_{ik} p_r \mathbb{D}_{jr} + \rho \, p_i \mathbb{D}_{ij} \mathbb{D}_{kr} p_r ) + \p_j(- \rho W_{ik} \psi'_{ij} )= 0\\
\delta W_{ij}: \quad & - \rho \psi'_{ij} - \p_t A_{ij}
+ \rho \psi'_{im} \p_m \lambda_j + \rho \psi''_{knij} W_{km} \p_n \lambda_m - \p_k ( p_r \mathbb{D}_{rk} A_{ij} )  + \p_k ( p_r \mathbb{D}_{rj}) A_{ik} = 0\\
\delta \rho: \quad & - p_i \mathbb{D}_{ik} \p_k \theta + p_i \mathbb{D}_{ik} \p_t \lambda_k + \frac{1}{2 \rho}  p_i p_j \mathbb{D}_{ij} - \psi - \p_t \theta + W_{ki} \psi'_{kj} \p_j \lambda_i = 0.
\end{aligned}
\end{equation}
With the definition
\begin{equation}\label{eq:def_v_1}
    v_k := \mathbb{D}_{kj} p_j = \mathbb{D}_{jk} p_j
\end{equation}
the first three equations of the system \eqref{eq:E-L_1} are identical to the system \eqref{eq:FDM} for $f_i = 0$.

As a second alternative, consider the action
\begin{equation}\label{eq:mixed_action}
    \begin{aligned}
    S_2[A, W, \rho, \theta, \lambda; \alpha, V] = \int_{\Omega \times [0,T]} dt d^3x & -\frac{1}{2} \frac{p_k p_k}{\rho} - \rho \psi(W) \\
    & + \frac{1}{\rho} (p_i p_j \p_j \lambda_i) + A_{ij}[ \p_t W_{ij} - (\alpha \times V)_{ij} - \p_j f_i ]\\
    & + \theta \p_t \rho + \lambda_i [ \p_j (-\rho W_{ki} \psi'_{kj} )],
    \end{aligned}
\end{equation}
where again
\begin{equation}\label{eq:def_p}
p_k := -[ A_{ij} \p_k W_{ij} - \p_j (A_{ij} W_{ik}) - \rho \p_k \theta + \rho \p_t \lambda_k ],
\end{equation}
and the fields $\alpha_{ij}$, $V_i$, and $f_i$ are considered fields that are not varied. Then its first variation, assuming all variations vanish on the boundary of $\Omega \times [0,T]$, is given by
\begin{equation}\label{eq:first_var_1}
\begin{aligned}
\delta S_2 = \int_{\Omega \times [0,T]} dt d^3x \  & \frac{1}{2 \rho^2} p_k p_k \delta \rho - \psi \delta \rho - \rho \psi'_{ij} \delta W_{ij} - \frac{1}{\rho^2} p_i p_j \p_j \lambda_i \delta \rho - \p_j \left(\frac{1}{\rho} p_i p_j\right) \delta \lambda_i \\
& + \delta A_{ij} [ \p_t W_{ij} - (\alpha \times V)_{ij}] - \p_t A_{ij} \delta W_{ij} \\
& + \delta \theta \p_t \rho - \p_t \theta \delta \rho + \delta \lambda_i  \p_j (- \rho W_{ki} \psi'_{kj}) \\
& + \p_j \lambda_i \left[ \delta \rho W_{ki} \psi'_{kj} + \rho \delta W_{ki} \psi'_{kj} + \rho W_{ki} \psi''_{kjmn} \delta W_{mn}\right]\\
& \qquad \qquad \qquad + \\
& {\color{blue} - \frac{1}{\rho} p_k \delta p_k + \frac{\p_j\lambda_i}{\rho} [p_i \delta p_j + p_j \delta p_i] }.
\end{aligned}
\end{equation}
Defining
\begin{equation}\label{eq:def_R}
R_k := -\rho^{-1} [ p_i \p_k \lambda_i + p_j \p_j \lambda_k - p_k ],
\end{equation}
the contribution to the first variation from the terms in blue in \eqref{eq:first_var_1} become
\begin{equation}\label{eq:first_var_2}
\begin{aligned}
\int_{\Omega \times [0,T]} dt d^3x & \ R_k \p_k W_{ij} \delta A_{ij} - \p_k (R_k A_{ij}) \delta W_{ij} + W_{ik} \p_j R_k \delta A_{ij} + (A_{ij} \p_j R_k) \delta W_{ik}\\
& - R_k \p_k \theta \delta \rho + \p_k (\rho R_k) \delta \theta + R_k \p_t \lambda_k \delta \rho - \p_t (\rho R_k) \delta \lambda_k.
\end{aligned}
\end{equation}
Thus, \eqref{eq:first_var_1} and \eqref{eq:first_var_2}, using the definitions \eqref{eq:def_p} and \eqref{eq:def_R}, imply the Euler-Lagrange equations,
\begin{equation}\label{eq:E-L}
\begin{aligned}
\delta A_{ij}: \quad & \p_t W_{ij} - (\alpha \times V)_{ij} - \p_j f_i + R_k \p_k W_{ij} + W_{ik} \p_j R_k = 0\\
\delta \theta: \quad & \p_t \rho + \p_k (\rho R_k) = 0\\
\delta \lambda_i : \quad & - \p_j \left(\rho^{-1} p_i p_j \right) + \p_j (- \rho W_{ki} \psi'_{kj}) - \p_t (\rho R_i) = 0\\
\delta W_{ij}: \quad & - \rho \psi'_{ij} - \p_t A_{ij} 
+ \rho \psi'_{ik} \p_k \lambda_j + \rho \psi''_{knij} W_{kp} \p_n \lambda_p - \p_k (R_k A_{ij}) + A_{ik} \p_k R_j = 0\\
\delta \rho: \quad & \frac{1}{2 \rho^2} p_k p_k - \psi - \frac{1}{\rho^2} p_i p_j \p_j \lambda_i - \p_t \theta + W_{ki} \psi'_{kj} \p_j \lambda_i - R_k \p_k \theta + R_k \p_t \lambda_k = 0.
\end{aligned}
\end{equation}
With the definition
\begin{equation}\label{eq:def_v}
    v_k := \frac{p_k}{\rho}
\end{equation}
the first three equations of the system \eqref{eq:E-L} may be written as 
\begin{equation}\label{eq:E-L_to_disloc_mech}
  \begin{aligned}
   \p_t W_{ij} + v_k \p_k W_{ij} + W_{ik} \p_j v_k  - (\alpha \times V)_{ij} - \p_j f_i & = W_{ik} \p_j ( v_i \p_k \lambda_i + v_j \p_j \lambda_k) + ( v_i \p_k \lambda_i + v_j \p_j \lambda_k) \p_k W_{ij} \\
  \p_t \rho + \p_k(\rho v_k) & = \p_k(\rho( v_i \p_k \lambda_i + v_j \p_j \lambda_k) )\\
  -\p_t(\rho v_i) - \p_j (\rho v_i v_j) + \p_j (- \rho W_{ki} \psi'_{kj}) & = \p_t (\rho( v_k \p_i \lambda_k + v_j \p_j \lambda_i) ).
  \end{aligned}  
\end{equation}
With the definitions \eqref{eq:def_p} and \eqref{eq:def_v} in force, one solution of the system \eqref{eq:E-L} can be generated by requiring that the fields $\lambda_i$ satisfy
\begin{equation}\label{eq:lambda_constr}
v_i \p_k \lambda_i + v_j \p_j \lambda_k = 0.
\end{equation}
Equations \eqref{eq:E-L} with the definitions \eqref{eq:def_p}, \eqref{eq:def_R} constitute $9+1+3+9+1+3+3 = 29$ equations in the $32$ variables $(A_{ij}, \theta, \lambda_i, W_{ij}, \rho, p_k, R_k, f_i)$ (the count can be reduced to 23 equations in 26 variables); adding equation \eqref{eq:lambda_constr} (with \eqref{eq:def_v} enforced) to the set of equations gives equal number of equations and unknowns. Thus, the system is not formally overdetermined.

With \eqref{eq:lambda_constr} enforced, a solution to the system \eqref{eq:E-L} with the definitions \eqref{eq:def_v}, \eqref{eq:def_R} is a solution  of \eqref{eq:FDM}, \textit{with $f_i$ determined}, in general. This is not entirely satisfactory either as, e.g., when doing nonlinear elasticity without defects, \eqref{eq:FDM} should hold with $\p_j f_i = 0$ and $\alpha_{ij} = 0$.
\section{The primal actions and their reduced state space}\label{sec:primal_action}
While not strictly necessary for the main goal of this work, namely, defining variational principles corresponding to the system \eqref{eq:FDM} in well-defined senses, motivation is provided here on how the actions \eqref{eq:mixed_action_1} and \eqref{eq:mixed_action} were arrived at. This will also be useful later for considering variations on the theme in Sec.~\ref{sec:obs}.

The main idea is to invoke a Legendre transform based change of variables and then consider the variational principle in a \textit{reduced state space}. 

 Consider
\begin{equation}\label{eq:primal_action}
\begin{aligned}
\widehat{S} = \int_{\Omega \times [0,T]} dt d^3x \ & \frac{1}{2} \rho v_i v_i - \rho \psi(W) \\
& + A_{ij} \left[ \p_t W_{ij} + v_k \p_k W_{ij} + W_{ik} \p_j v_k - (\alpha \times V)_{ij} - \p_j f_i \right] \\
& + \theta [ \p_t \rho + \p_i(\rho v_i)] \\
& + \lambda_i \left[ \p_j ( - \rho W_{ki} \psi'_{kj}) - \p_t (\rho v_i) - \p_j (\rho v_i v_j) \right],
\end{aligned}
\end{equation}
where the equations of \eqref{eq:FDM} have been imposed with Lagrange multipliers along with the usual, customary choice in mechanics of the difference of kinetic energy and potential energy. 

Integrate by parts in \eqref{eq:primal_action} to expose linear terms in $v_i$, assuming Lagrange multipliers vanish on the boundary of the space-time domain. Then
\begin{equation}
    \label{eq:motiv_p}
    \begin{aligned}
    \widehat{S} = \int_{\Omega \times [0,T]} dt d^3x \ &  \frac{1}{2} \rho v_i v_i - \rho \psi(W) \\
    & + [ A_{ij} \p_k W_{ij} - \p_j(A_{ij} W_{ik}) - \rho \p_k \theta + \rho \p_t \lambda_k ] v_k \\
    & + \rho v_i v_j \p_j \lambda_i \\
    & + A_{ij} [ \p_t W_{ij} - (\alpha \times V)_{ij} - \p_j f_i]\\
    & + \theta \p_t \rho + \lambda_i [ \p_j (-\rho W_{ki} \psi'_{kj} )].
    \end{aligned}
\end{equation}
Define $K(v) = \frac{1}{2} \rho v_i v_i$, which is convex in $v$ and therefore $K'_i : = \p_{v_i} K = \rho v_i$ is an invertible function on the space of spatial vectors. Suppose further that we consider the following reduced state space defined by eliminating $v_i$ in terms of the rest of the fields appearing in \eqref{eq:def_p}:
\begin{equation*}
\begin{aligned}
p_k & := -[ A_{ij} \p_k W_{ij} - \p_j (A_{ij} W_{ik}) - \rho \p_k \theta + \rho \p_t \lambda_k ]\\
v_i(p) & := \left(K^{'-1}\right)_i (p) = \frac{p_i}{\rho}.
\end{aligned}
\end{equation*}
Then, defining the function $K^*$ given by
\[
K^*(p) := p_i v_i(p) - K(v(p)) = \frac{1}{2} \frac{p_i p_i}{\rho}
\]
\eqref{eq:motiv_p} becomes
\begin{equation}
    \begin{aligned}
    S_2[A, W, \rho, \theta, \lambda; \alpha, V, f] = \int_{\Omega \times [0,T]} dt d^3x & -K^*(p) - \rho \psi(W) \\
    & + \frac{1}{\rho} (p_i p_j \p_j \lambda_i) + A_{ij}[ \p_t W_{ij} - (\alpha \times V)_{ij} - \p_j f_i ]\\
    & + \theta \p_t \rho + \lambda_i [ \p_j (-\rho W_{ki} \psi'_{kj} )].
    \end{aligned}
\end{equation}
which is the action \eqref{eq:mixed_action}.

If instead now we define 
\[
M(v) := \frac{1}{2} \rho v_i v_i + v_i \rho \, \p_j \lambda_i v_j,
\]
and consider the reduced state space defined by eliminating $v_i$ in terms of the rest of the fields appearing in \eqref{eq:def_p_1}:
\begin{equation*}
\begin{aligned}
p_k & := -[ A_{ij} \p_k W_{ij} - \p_j (A_{ij} W_{ik}) - \rho \p_k \theta + \rho \p_t \lambda_k ]\\
v_i(p) & := \mathbb{D}_{ij} p_j
\end{aligned}
\end{equation*}
\textit{which assumes that} $\mathbb{D} = \mathbb{L}^{-1}$ \textit{always exists}. Furthermore, defining a function $M^*$ as
\[
 M^*(p):= p_i v_i(p) - M(v(p)) = \frac{1}{2} p_i \mathbb{D}_{ij} p_j,
\]
\eqref{eq:motiv_p} becomes
\begin{equation}
    \begin{aligned}
    S_1[A, W, \rho, \theta, \lambda; \alpha, V] = \int_{\Omega \times [0,T]} dt d^3x & -M^*(p) - \rho \psi(W) \\
    & + A_{ij}[ \p_t W_{ij} - (\alpha \times V)_{ij} + \p_j f_i ]\\
    & + \theta \p_t \rho + \lambda_i [ \p_j (-\rho W_{ki} \psi'_{kj} )].
    \end{aligned}
\end{equation}
which is the action \eqref{eq:mixed_action_1} when $f_i = 0$ from the outset.

Of course, the considerations in this Section simply outline a pathway/motivation for generating the actions \eqref{eq:mixed_action_1} and \eqref{eq:mixed_action} whose Euler-Lagrange equations have a desired property, and hence they do not require the vanishing of the Lagrange Multiplier fields on the boundary of $\Omega \times [0,T]$.

We note here the following feature of our actions:
\begin{itemize}

    \item We require action functionals which contain derivatives of their constituent fields in the action density in `more than linear' combinations. With a target set of field equations as desired E-L equations in mind, this cannot be achieved simply by imposing the desired field equations with Lagrange Multiplier fields (the Lagrange multiplier fields would only appear linearly). This leads to considering additional convex potentials in some of the basic fields of the desired target equations and then trying to eliminate these basic fields in terms of the Lagrange multiplier fields. But it is not directly obvious that the combination of a) the addition of the potentials and b) the elimination of some of the basic fields through the adopted change of variables does not interfere with recovering the target set of equations as the E-L equations of the developed action - even though at the starting point the target set was accommodated by Lagrange Multiplier fields. In fact, our action $S_2$ shows that this may not always be possible for a given target. This issue is discussed further in observation \ref{itm:obs_8} in Sec.~\ref{sec:obs}.
    
\end{itemize}

\section{Contact with fracton models: an action for geometrically linear dislocation-disclination mechanics in $\mathbf{3+1}$-D}\label{sec:fracton}
With reference to the classical elastic theory of defects and its fields \cite{dewit1971relation, dewit1973theory, dewit1973theory4, kroner1981continuum} described in the Appendix \ref{app:app} (we provide a perhaps fresh perspective on the meaning of the `plastic' fields of DeWit), here we start with the primal action as motivation and deduce the proposed `dual' action for dislocation-disclination mechanics $(3+1)$-d, showing convergence with current research trends in fracton-elasticity duality \cite{pretko2018fracton, gromov2020duality} in $(2+1)$-d. In what follows $\mathbb{C}$ is fourth-order tensor of elastic moduli with major and minor symmetries, $\veps$ is the symmetric part of the elastic distortion (not necessarily a symmetrized gradient), and $v$ is the material velocity field. We also employ the notation defined in \eqref{eq:not_(a)sym}.

The target field equations for the geometrically linear defect theory are related to the system \eqref{eq:FDM} as follows: we consider an elastic distortion about the identity of the form $\delta_{ij} + u_{ij}$ when its inverse $W_{ij} \sim \delta_{ij} - u_{ij}$ (which is appropriate for `small' $u_{ij}$), drop all explicit nonlinear terms in the system \eqref{eq:FDM} and, since $u_{ij}$ is small, it can be shown that the stress term $(-\rho W_{ki} \psi'_{kj})$ in \eqref{eq:blm} can be written as $\mathbb{C}_{ijkl} \varepsilon_{kl}$, when expanded about the state $W_{ij} = \delta_{ij}$, assuming in that state the stress vanishes. Here we consider defects beyond dislocations as described in the Appendix \ref{app:app}, so the $J_{ij} = \alpha^* \times V$ replaces the flux $\alpha \times V$ in the second line of \eqref{eq:sd_primal}; this line imposes the `linearized' version of \eqref{eq:v-W_comp} in the above sense. The third line incorporates the fundamental statement of incompatibility of elastic strain sourced by the dislocation and disclination fields; the fourth line is the linearized version of \eqref{eq:blm}. Under the above ansatz, \eqref{eq:bom} simply implies that $\p_t \rho \sim 0$ and in this setting $\rho$ is assumed a specified field, constant in time.

Consider
\begin{equation}
    \label{eq:sd_primal}
    \begin{aligned}
    \widehat{S} = \int_{\Omega \times [0,T]} dt d^3x \ & \frac{1}{2} \rho v_i v_i  - \frac{1}{2} \veps_{ij} \mathbb{C}_{ijkl} \veps_{kl} \\
    & + A_{ij} [ \p_j v_i - \p_t u_{ij} - J_{ij} ]\\
    & + \gamma_{rp} [ e_{rqi} e_{pkj} \p_q \p_k \veps_{ij} - s_{rp}]\\
    & + \lambda_i [ \p_j (\mathbb{C}_{ijkl} \veps_{kl}) - \rho \p_t v_i],
    \end{aligned}
\end{equation}
where $A_{ij}, \gamma_{rp}, \lambda_i$ are Lagrange multiplier fields, and $J_{ij}$ and $\rho$ are assumed to be given fields over the space-time domain $\Omega \times [0,T]$. Exposing linear terms in $v_i$ and $\veps_{ij}$
\begin{equation}\label{eq:sd_motiv_p}
    \begin{aligned}
    \widehat{S} = \int_{\Omega \times [0,T]} dt d^3x \ & \frac{1}{2} \rho v_i v_i  - \frac{1}{2} \veps_{ij} \mathbb{C}_{ijkl} \veps_{kl} \\
    & + [\p_t(\rho \lambda_i) - \p_j A_{ij}] v_i\\
    & + [ \p_t \overline{A}_{ij} + e_{iqr} e_{jkp} \p_q \p_k \gamma_{rp} - \mathbb{C}_{ijkl} \p_l \lambda_k ]\veps_{ij}\\
    & + \p_t \widetilde{A}_{ij} r_{ij} - A_{ij} J_{ij} - \gamma_{rp} s_{rp},
    \end{aligned}
\end{equation}
where the Lagrange multipliers have been assumed to vanish on the boundary of the space-time domain. Define the convex functions $K(v)$ and $U(\veps)$ of their respective arguments by
\[
\begin{aligned}
K(v) &:= \frac{1}{2} \rho v_i v_i \\
U(\veps) &:= \frac{1}{2} \veps_{ij} \mathbb{C}_{ijkl} \veps_{kl},
\end{aligned}
\]
with $\mathbb{C}_{ijkl}$ is assumed to be positive definite on the space of symmetric second-order tensors.  Consider now the elimination of $v_i$ and $\veps_{ij}$ in terms of the rest of the fields through
\[
\begin{aligned}
p_i & := -[\p_t (\rho \lambda_i) - \p_j A_{ij}]\\
v_i(p) & := \left(K^{'-1}\right)_i (p) = \frac{p_i}{\rho}\\
\sigma_{ij} & := \p_t \overline{A}_{ij} + e_{iqr} e_{jkp} \p_q \p_k \gamma_{rp} - \mathbb{C}_{ijkl} \p_l \lambda_k\\
\veps_{ij} (\sigma) & := \left( U^{'-1} \right)_{ij} (\sigma) = \mathbb{S}_{ijkl} \sigma_{kl}
\end{aligned}
\]
where $\mathbb{S}$ is the positive definite tensor of \textit{elastic compliance}, with $\mathbb{S} = \mathbb{C}^{-1}$ on the space of symmetric second order tensors. Then, invoking the Legendre transforms of $K$ and $U$ given by
\begin{equation}
    \begin{aligned}
    K^*(p) & = p_i v_i(p) - K(v(p)) = \frac{1}{2} \frac{p_i p_i}{\rho}\\
    U^*(\sigma) & = \sigma_{ij} \veps_{ij}(\sigma) - U(\veps(\sigma)) = \frac{1}{2} \sigma_{ij} \mathbb{S}_{ijkl} \sigma_{kl}
    \end{aligned}
\end{equation}
the proposed \textit{ `dual' action} for geometrically linear dislocation-disclination mechanics is
\begin{equation}\label{eq:sd_dual_action}
    \begin{aligned}
    S[A,\lambda,\gamma,r;\rho,J,s] := \int_{\Omega \times [0,T]} dt d^3x \ - K^*(p) + U^*(\sigma) + r_{ij} \p_t \widetilde{A}_{ij}  - A_{ij} J_{ij} - \gamma_{rp} s_{rp}.
    \end{aligned}
\end{equation}

For variations that vanish on the boundary of the space-time domain, the first variation of the dual action in \eqref{eq:sd_dual_action} is given by
\begin{equation*}
    \begin{aligned}
    \delta S = \int_{\Omega \times [0,T]} dt d^3x \ & - \frac{p_i}{\rho} [\p_j \delta A_{ij} - \p_t (\rho \delta \lambda_i)]\\
    & + \mathbb{S}_{ijmn} \sigma_{mn} [ \p_t \delta \overline{A}_{ij} + e_{iqr} e_{jkp} \p_q \p_k \delta \gamma_{rp} - \mathbb{C}_{ijkl} \p_l \delta \lambda_k]\\
    & - \delta \widetilde{A}_{ij} \p_t r_{ij} + \delta r_{ij} \p_t \widetilde{A}_{ij} - J_{ij} \delta A_{ij} - s_{rp} \delta{\gamma}_{rp}
    \end{aligned}
\end{equation*}
yielding the Euler-Lagrange equations
\begin{equation*}
    \begin{aligned}
    \delta A_{ij}: \ & \p_j v_i - \p_t ( \veps_{ij} + r_{ij}) - J_{ij} = 0\\
    \delta \lambda_i: \ & - \rho \p_t v_i + \p_j (\mathbb{C}_{ijkl} \veps_{kl}) = 0 \\
    \delta \gamma_{rp}: \ & e_{rqi} e_{pkj} \p_k \p_q \veps_{ij} - s_{rp} = 0\\
    \delta r_{ij}: \ & \p_t \widetilde{A}_{ij} = 0.
    \end{aligned}
\end{equation*}
\section{Discussion}\label{sec:obs}
Some observations about, and implications of, the developed framework are discussed. The remarks are made in the context of the action $S_1$ but they apply to the action $S_2$ as well.
\begin{enumerate}
    \item \label{itm:obs_1} When $V_i$ and $\alpha_{ij}$ are assumed as specified functions of space and time (as assumed in the development above) the Euler-Lagrange equations \eqref{eq:E-L_1} with the definition \eqref{eq:def_v_1} amount to those of the nonlinear elastic theory of dislocations, reducing to nonlinear elasticity when $\alpha_{ij} = 0$, as shown in Sec. \ref{sec:FDM}, as already mentioned.
    
    It can be checked that when $\alpha$ is constrained as $\alpha = - \mbox{curl}\, W$, and $V$ is constrained through a constitutive equation in terms of $W, \mbox{curl}\,W, \mbox{curl}\,\mbox{curl}\,W$, the E-L equations corresponding to variations ($\delta A_{ij}, \delta \theta, \delta \lambda_k$) remain unchanged (with the obvious substitutions of $\alpha = - \mbox{curl}\, W$ and $V = V(\mbox{curl}\, W, W, \mbox{curl}\,\mbox{curl}\, W)$) and the E-L equation corresponding to $\delta W_{ij}$ is what sees substantial change.
    
    In this connection, it is interesting to note that for a particular class of such constitutive assumptions, the presented framework embeds a strongly dissipative, out-of-equilibrium system within a variational principle.
    
    An exactly similar observation pertains to the inclusion of an argument of $\mbox{curl}\, W$ in $\psi$ (reflecting the physics of including a core energy), with appropriate changes in the functional forms of the Cauchy stress in \eqref{eq:blm} and the dislocation velocity in \eqref{eq:v-W_comp} following the dictates of second law of thermodynamics (restricted here to mechanical processes) and non-negative dissipation \cite{acharya2011microcanonical}.
    
    \item  Due to invariance under superposed rigid motions of the energy density $\psi(W)$, it can depend on $W$ only through the combination $\mathcal{B} = WW^T$, say $\psi(W) = \hat{\psi}(\mathcal{B}(W))$. Then, since $\p_{W_{ij}} \psi = 2 W_{lj} \p_{\mathcal{B}_{il}} \hat{\psi}$, $\p_{W_{ij}} \psi =  0$ implies $\p_{\mathcal{B}_{il}} \hat{\psi} = 0$ as $W$ is assumed invertible. Since it is a physically natural property of any elastic response function that when the inverse elastic distortion $W$ is \textit{any} orthogonal tensor (and therefore the elastic distortion as well), $\p_{\mathcal{B}_{il}} \hat{\psi}$ evaluates to zero, this implies that the function $W \mapsto \psi'(W)$ is not invertible and hence a Legendre transform cannot be invoked for it. Furthermore, in the context of crystal elasticity, the function $\hat{\psi}(\mathcal{B})$ cannot be convex to reflect lattice-periodicity, i.e. the existence of non-trivial homogeneous deformations that nevertheless leave the lattice, and hence its energy density, invariant, and therefore the function $W \mapsto \psi'(W)$ is again not one-to-one.
    
    \item Linearizing the first two terms in the expression for $p_k$ about a state $(A_{ij}, W_{ij})$ in \eqref{eq:def_p} one obtains
    \[
    \Delta p_k \sim -[ A_{ij} (\p_k \Delta W_{ij} - \p_j \Delta W_{ik}) + (\p_k W_{ij} - \p_j W_{ik}) \Delta A_{ij} - \Delta W_{ik} \p_j A_{ij} -\p_j A_{ij} \Delta W_{ij}].
    \]
    A quadratic expression in $\Delta p_k$ approximating its analogous term in the action \eqref{eq:mixed_action} while considering only the first two terms in the above expression bears some similarity with the spatial part of the postulated minimal coupling Lagrangian of \cite[Eqn. (115)]{beekman2017dual}, $\mathcal{L}_{\mbox{min.coup.}}$. The potential utility of this analogy coupled with the physically mandated multi-well nonconvexity of the energy density $\hat{\psi}(\mathcal{B}(W))$ modeling the postulated Higgs potential of \cite[Eqn. (112)]{beekman2017dual} is an important direction for future work.
    
    \item The imposition of the fundamental compatibility relation \eqref{eq:v-W_comp} between the inverse elastic distortion, the velocity gradient, and the plastic distortion rate produced due to dislocation motion (\cite[Sec.~5.3]{acharya2015dislocation}-\cite[Appendix B]{acharya2011microcanonical}) with a Lagrange multiplier field naturally gives rise to a `Kalb-Ramond'-like Lagrangian \cite{garcia2018multipole, beekman2017dual, kalb1974classical} given by $A_{ij} (\alpha \times V)_{ij}$ in the action \eqref{eq:mixed_action_1} (with the skew-symmetric pair of indices of the Kalb-Ramond field associated with 2-vectors on surfaces dualized to one index associated with the normal to the 2-vector surface element in the usual way).
    
    \item The variational formulation embeds the FDM system for nonlinear dislocation dynamics within a larger system of pde given by \eqref{eq:E-L_1}. Furthermore, it is interesting to note that this is in fact achieved even if the appearance of the function $\psi(W)$ on the first line of the Lagrangian in  \eqref{eq:mixed_action} is replaced by any \textit{arbitrary} smooth function, say $\mathcal{F}$, of the same argument. It seems not unreasonable to expect that these two features taken together can be of some help in facilitating the existence of solutions to the smaller FDM system. It is interesting that the E-L equations \eqref{eq:E-L_1} require solutions of the FDM system (interpreted in terms of \eqref{eq:def_v_1} and \eqref{eq:def_p_1}) to satisfy more differential relations (\eqref{eq:E-L_1}$_{5,6}$) with other fields, but without overcontraints. 
    
    \item In a completely formal sense, ignoring the terms $-\p_t(\rho v_i) - \p_j(\rho v_i v_j)$  on the last line of \eqref{eq:primal_action} and following through with its consequences delivers the action principle corresponding to quasi-static FDM.
    
    \item In the context of the strict goal of deriving an action principle whose E-L equations contain the FDM system, it is clear from our considerations that the occurrence of $\psi(W)$ on the first line of the action in \eqref{eq:mixed_action} can be replaced by any smooth function, say $\mathcal{F}(W)$, with impunity, as already observed. In fact, it seems reasonable to explore replacing both the kinetic and potential energy terms on the first line of \eqref{eq:mixed_action} by convex functions of $v$ and $W$, respectively, to see if the equations of FDM, with appropriate interpretation, can be recovered for arbitrary convex functions beyond quadratic dependence. The consequences of this degree of generality, and how it may be exploited, is a direction for future work. Some progress in answering this question has been made in \cite{acharya2021variational}. Here, in Sec. \ref{sec:dual_elast}, we show the derivation of a family of variational principles whose E-L equations are the field equations of nonlinear elastostatics written on a reference configuration for an elastic material whose (first Piola-Kirchhoff) stress response is not necessarily hyperelastic.
    
    \item \label{itm:obs_8} Based on the experience with the action $S_2$, it seems important to understand the previous question in the context of the choice $f_i = 0$. A conjecture in this regard is that if a Legendre transform motivated change of variables is invoked on some `primal' field, say $v \to p$, it must be such that after transformation the transformed variable ($p$) should not appear in the dual action except as an argument of the introduced dual potential (e.g., this did not happen in $S_2$, but did in $S_1$). A (dis)proof of this conjecture would be desirable. If true, then this principle can guide the choice of the admissible class of convex potentials that can be admitted (for a given specific primal field), which depends crucially on the structure of the target pde system (as demonstrated by the choices $K$ and $M$ in Sec.~\ref{sec:primal_action}), another desirable feature.
\end{enumerate}

Finally, we note that the `Coulomb-nematic' phase of \cite{zaanen2004duality} involving an order parameter with anti-parallel Burgers vector everywhere appears to be rather relevant to a description of macroscopic plasticity. It could be useful to understand the relation of such an order parameter to Kroupa's \cite{kroupa1962continuous} loop density and to what extent the EFT describes its dynamics, which would necessarily have to include a description of work-hardening. This can be beneficial for the study of plasticity via EFT, adapting the treatments of \cite{beekman2017dual, beekman2017dual1}.

\subsection{Dual variational principles for nonlinear elastostatics}\label{sec:dual_elast}
Consider the field equations
\begin{equation}\label{eq:nonlin_elast}
\begin{aligned}
0 & = \p_j K_{ij}(F) \\
F_{ij} &= \p_j y_i,
\end{aligned}
\end{equation}
where $K$ is a given tensor valued function of invertible tensors $F$ (delivering the First Piola-Kirchhoff stress tensor for a prescribed deformation gradient), and the derivatives are now w.r.t. rectangular Cartesian coordinates on a fixed reference configuration $\Omega_R$ (\textit{$K$ is not assumed to necessarily be a gradient of a scalar valued function on the space of invertible tensors}). The only restrictions on $K$ we require are that it be sufficiently smooth in its argument with $\p_F K$ and $\p_{FF} K$ being bounded, and that it be of the form 
\[
K(F) = (\mbox{det} F)\, F \,S\!\left(F^TF \right)
\]
for $S$ being any arbitrary symmetric tensor valued function of a symmetric tensor; this allows frame-indifference to be satisfied.

We take the inner-product of the equations with Lagrange multiplier fields $(\lambda, A)$ that vanish on the boundary $\p \Omega_R$, and introduce the notation 
\[
U := (y, F).
\]
A key step, and an assumption implicit in the procedure for defining the `dual' functional $S_E[A, \lambda]$ (see \eqref{eq:dual_vp_elast} below), is to invoke a scalar valued function $H(U)$ such that, for the auxiliary function $M(U)$ defined as
\begin{subequations}\label{eq:M}
\begin{align}
M(U, \nabla \lambda) &:= H(U) -  K_{ij}(F(U)) \p_j \lambda_i, \\
(\p_{y} M, \quad \p_F M) =: \p_U M (U, \nabla \lambda) &= P; \quad (\p_U M)_{i, kl} =  (\p_{y_i} H, \qquad \p_{F_{kl}} H - (\p_{F_{kl}} K_{ij}) \p_j \lambda_i),  \label{eq:U-P}
\end{align}
\end{subequations}
\eqref{eq:U-P} has a solution $U(P, \nabla \lambda)$ for prescribed $P$ (where the dimension of the arrays $U$ and $P$ are obviously the same). Essentially, the goal of the introduction of $H(U)$ is to ensure that \eqref{eq:U-P} is solvable for $U$; indeed if $H$ could be chosen so as to make $M$ convex in $U$, then this would be guaranteed\footnote{For the action functional $S_1$ for dislocation mechanics considered in Secs. \ref{sec:main_action} and \ref{sec:primal_action}, the kinetic energy density plays the role of the convex $H$ (in the velocity field $v$).}.

Consider the auxiliary functional
\begin{equation*}
    \widehat{S}_E [y,F, \lambda, A] = \int_{\Omega_R} d^3x \ H(y,F) - K_{ij}\p_j \lambda_i  (F) +  y_i \p_j A_{ij} + F_{ij} A_{ij} = \int_{\Omega_R} d^3x \  M(U, \nabla \lambda) - U \cdot P
\end{equation*}
where
\[
P := - (\mbox{div} A, A).
\]
Now defining
\begin{equation} \label{eq:M*}
M^*(P, \nabla \lambda) : = P \cdot U(P, \nabla \lambda) - M(U(P, \nabla \lambda), \nabla \lambda)
\end{equation}
it can be verified using \eqref{eq:M} that
\begin{equation}\label{eq:M*deriv}
\p_P M^* = U.
\end{equation}

Consider now the dual functional
\begin{equation}\label{eq:dual_vp_elast}
S_E[A, \lambda] := \int_{\Omega_R} d^3x \ - M^*(P, \nabla \lambda).
\end{equation}
Its first variation is given by
\[
\delta S_E = \int_{\Omega_R} d^3x \  - \p_P M^* \cdot \delta P - (\p_{\nabla \lambda} M^*)_{ij} \p_j \delta \lambda_i.
\]
Now, \eqref{eq:M*} implies, using \eqref{eq:M}, that
\begin{equation*}
   \p_{\nabla \lambda} M^* = P \cdot \p_{\nabla \lambda} U - \p_U M \cdot \p_{\nabla \lambda} U - \p_{\nabla \lambda} M =  K
\end{equation*}
and noting \eqref{eq:M*deriv}, we have
\[
\delta S =  \int_{\Omega_R} d^3x \ \delta A_{ij} ( F_{ij} - \p_j y_i ) + \delta \lambda_i \p_j K_{ij}, 
\]
\textit{which implies the E-L equations \eqref{eq:nonlin_elast}}. Of course, here the fields $F, y$, and hence $K(F)$, are mappings on the reference configuration through the fields $(P, \nabla \lambda)$.

We note that \eqref{eq:dual_vp_elast} defines a family of variational principles for nonlinear elastostatics parametrized by the choice of the function $H$ in the definition of the potential $M$ in \eqref{eq:M}.

As an example, consider the specific choice
\begin{equation*}
    H(y, F)  = \frac{1}{2} \mu y_i y_i + \frac{1}{2} \beta F_{ij}F_{ij},
\end{equation*}
where $\mu, \beta$ are positive scalar constants which can be arbitrarily specified. Then $M(U, \nabla \lambda) = H(U) - K_{ij}(F) \p_j \lambda_i$, and
\begin{equation*}
    \begin{aligned}
    \p_{y_i} M & = \mu y_i \\
    \p_{F_{ij}} M &= \beta F_{ij} - (\p_F K)_{rsij} (\nabla \lambda)_{rs}.
    \end{aligned}
\end{equation*}
 $P = \p_U M$ implies
\begin{equation}\label{eq:mapping}
    \begin{aligned}
    y_i(P, \nabla \lambda) & = - \frac{1}{\mu} \p_k A_{ik} \\
    (\p_{F} M)_{ij} = - A_{ij} \Longrightarrow F_{ij} (P, \nabla \lambda) & = \frac{1}{\beta} \left( - A_{ij} + (\p_F K)_{rsij} (\nabla \lambda)_{rs} \right);
    \end{aligned}
\end{equation}
in the following, we will abuse notation to write $F(A, \nabla \lambda)$ as well. Thus, the governing (E-L) equations of the dual problem in the fields $(\lambda, A)$ are
\begin{subequations}\label{eq:gov_dual}
    \begin{align}
    \frac{1}{\mu} \p_j \p_k A_{ik} + F_{ij}(A, \nabla \lambda) & = 0 \label{eq:comp}\\
    \p_j K_{ij} (F (A, \nabla \lambda)) & = 0. \label{eq:elast_equil}
    \end{align}
\end{subequations}

In order to get a rough sense of what may be involved in solving this system of equations, `parametrize' the field $A$ as
\[
A_{ij} (x) = \beta A^0_{ij} + H_{ij} (x)
\]
for $x \in \Omega_R$, and $A^0$ an arbitrary, constant invertible tensor. Viewing \eqref{eq:elast_equil} as the equation for $\lambda$, the existence of infinitesimal perturbations $\delta \lambda$ with continuous $\nabla \delta \lambda$ (given $A$), about a state $(\lambda, A)$ that satisfies equilibrium, is controlled by the properties of the matrix field $\p_{\nabla \lambda} K$, or the (strong) ellipticity of \eqref{eq:elast_equil}. Now, 
\[
F_{mn} (A, \nabla \lambda) = - A^0_{mn} - \frac{1}{\beta} H_{mn} + \frac{1}{\beta} \left[ (\p_F K) ( F(A, \nabla \lambda)) \right]_{pqmn} (\nabla \lambda)_{pq}
\]
implies
\begin{equation*}
    \begin{aligned}
    & \left( \delta_{im} \delta_{nj} - \frac{1}{\beta} (\nabla \lambda)_{pq} (\p_{F F} K)_{pqmnij} \right)  (\p_{\nabla \lambda} F)_{ijrs} = \frac{1}{\beta} (\p_F K)_{rsmn} \\
    & \left( \delta_{im} \delta_{nj} - \frac{1}{\beta} (\nabla \lambda)_{pq} (\p_{F F} K)_{pqmnij} \right)  (\p_A F)_{ijrs} = - \frac{1}{\beta}\delta_{mr} \delta_{ns}
    \end{aligned}
\end{equation*}
so that \textit{ for the parameter range $\beta \gg 1$}
\[
(\p_{\nabla \lambda} F)_{mnrs} \approx \frac{1}{\beta} (\p_F K)_{rsmn}
\]
is a good approximation, resulting in
\begin{equation*}
    (\p_{\nabla \lambda} K)_{ijrs} = (\p_{F} K)_{ijmn} (\p_{\nabla \lambda} F)_{mnrs} \approx  \frac{1}{\beta} (\p_F K)_{ijmn}(\p_F K)_{rsmn} \ge 0
\end{equation*}
in the sense of a quadratic form on second-order tensors. Thus, for $\beta \gg 1$, \textit{regardless of whether strong ellipticity fails for \eqref{eq:nonlin_elast} or not in solving for the deformation $y$, \eqref{eq:elast_equil} is expected to have better properties with regard to obtaining solutions for $\lambda$, and the extent of allowed nonuinqueness of incremental solutions out of general states}. We also have
\[
(\p_A K)_{ijrs} = (\p_{F} K)_{ijmn} (\p_A F)_{mnrs} \approx -\frac{1}{\beta} (\p_{F} K)_{ijrs}.
\]
 Since
\[
\p_j K_{ij} \approx  \frac{1}{\beta} (\p_F K)_{ijmn}(\p_F K)_{rsmn} \p_j \p_s \lambda_r - \frac{1}{\beta} (\p_F K)_{ijrs} \p_j H_{rs}
\]
for $\beta \gg 1$, a reasonable approximation to solutions of the quasilinear second-order system \eqref{eq:gov_dual} is expected to be provided by solutions to the \textit{linear, constant-coefficient} system
\begin{equation}
    \begin{aligned}
    \frac{1}{\mu} \p_j\p_k H_{ik} - \frac{1}{\beta} H_{ij} + \frac{1}{\beta} (\p_F K)^0_{pqij} \p_q \lambda_p& = A^0_{ij} \\
    \left[(\p_F K)^0_{ijmn}(\p_F K)^0_{rsmn}\right] \p_j \p_s \lambda_r - (\p_F K)^0_{ijrs} \p_j H_{rs} & =  0 \\
   - A^0_{ij} - \frac{1}{\beta} H_{ij} + \frac{1}{\beta} (\p_F K)^0_{pqij} \p_q \lambda_p   & = F_{ij} \\
   \beta A^0_{ij} + H_{ij} & = A_{ij}
    \end{aligned}
\end{equation}
where the superscript $^0$ represents an evaluation at $- A^0$. We note that \textit{$-A^0$ is arbitrary and may well be a state where $\p_F K$ is not strongly elliptic and corresponds to a `falling part of a stress-strain curve'}. 
 
The corresponding solution for $y_i$ of the primal problem is
\[
y_i = -\frac{1}{\mu} \p_k H_{ik}
\]
(and this satisfies the consistency condition that $\p_j y_i = F_{ij}$).

Higher-order corrections to this approximation seem to be computable, as well as solutions to the full system \eqref{eq:gov_dual}.

Dirichlet boundary conditions
\[
y_i = \bar{y}_i \qquad \mbox{on} \quad \p \Omega_r
\]
translate to the boundary condition
\[
\p_k H_{ik} = - \mu \bar{y}_i
\]
for the dual problem.

Solutions $(y,F)$ of \eqref{eq:gov_dual} are extremals of a well-behaved variational problem \eqref{eq:dual_vp_elast}  and define particular solutions, through the mapping \eqref{eq:mapping}, to the generally non-elliptic second-order system of pde \eqref{eq:nonlin_elast} of nonlinear elasticity that does not necessarily emanate from a variational principle. These solutions of \eqref{eq:nonlin_elast} can involve states where strong ellipticity is violated.

\section*{Acknowledgments}
This work was supported by the grant NSF OIA-DMR \#2021019. I thank Ira Rothstein and Shashin Pavaskar from whom I heard about the Kalb-Ramond construct, and for discussions.
\begin{appendix}
\section{Appendix: Geometrically linear dislocation-disclination defect theory in 3+1-D}\label{app:app}
For the geometrically linear model we consider displacements, $u_i$ from a fixed background domain in Euclidean ambient space and do not distinguish between material and spatial time derivatives. Both the displacement and velocity fields are allowed to develop terminating discontinuities on 2-d spatial surfaces that can evolve in time. Thus, $\p_j u_i$ and $\p_j v_i$ can both become singular on the surfaces of discontinuity and no longer remain integrable functions, but note that the $v_i$ remains integrable, even though possibly discontinuous. In the sense of distributions, $\p_j v_i$, is still a gradient, its singular part denoted by $v^{(s)}_{ij}$, concentrated on the surfaces of its discontinuity, is not necessarily curl-free, and we remove this singular part from $\p_j v_i$ to define the latter's regular part $v_{ij}$ as
\begin{equation}\label{eq:strn_rate_decomp}
\p_j v_i - v^{(s)}_{ij} =: v_{ij}.
\end{equation}
In the theory of plasticity, $v^{(s)}_{ij}$ is generalized to be an independent field not necessarily slaved to $\p_j v_i$ and completely determined by it - in this sense, it is an integrable function, perhaps with strong concentrations, which corresponds to a `zoomed-in' microscopic view, of the above macroscopic singular viewpoint. Similarly, the velocity field is continuous, without causing any loss of essential topological information and there being no essential problem with integration by parts. With this understanding, the statement \eqref{eq:strn_rate_decomp} is referred to as the decomposition of total velocity gradient into elastic (regular) and plastic (singular) parts.
In similar manner we consider a decomposition of the displacement gradient into regular and `singular' parts:
\begin{equation}
    \label{eq:displ_grad_decomp}
    \p_j u_i - u^{(s)}_{ij} =: u_{ij}
\end{equation}
The derivatives in \eqref{eq:strn_rate_decomp} and \eqref{eq:displ_grad_decomp} are in the sense of distributions so that their mixed-partial derivatives commute, and the relations
\[
\begin{aligned}
- e_{jrk} \p_r v^{(s)}_{ik} & = e_{jrk} \p_r v_{ik}\\
- e_{jrk} \p_r u^{(s)}_{ik} & = e_{jrk} \p_r u_{ik}
\end{aligned}
\]
hold.

To introduce disclinations the possibility of the regular part $u_{ij}$ of $\p_j u_i$ developing terminating discontinuities along surfaces is considered. In that case,
\begin{equation}
\label{eq:2-elastic_distortion_decomp}
\p_k u_{ij} - u^{(s)}_{ijk} =: u_{ijk}
\end{equation}
and one assumes that $u^{(s)}_{ijk}$ is skew int its first two indices, i.e. only the elastic rotation gradient can become singular and not the elastic strain gradient. In this case, the representation
\[
u^{(s)}_{ijk} = e_{ijp} \omega^{(s)}_{pk}
\]
holds. Of course, in the setting being considered there is nothing special about the assumption that only the elastic rotation gradient can become singular, and the notion of generalized disclinations can (and has been) introduced recently \cite{zhang2018relevance,zhang2018finite} where the entire elastic distortion (strain + rotation) gradient is allowed to develop terminating discontinuities. Here, we continue simply with the case of the classical disclination:
\[
\begin{aligned}
\p_k u_{ij} - e_{ijr} \omega^{(s)}_{rk} & = u_{ijk}\\
e_{ijr} \theta_{rm} &  = e_{mlk} \p_l u_{ijk} \qquad \mbox{where}
\qquad - e_{mlk} \p_l \omega^{(s)}_{rk} =: \theta_{rm}
\end{aligned}
\]
is the \textit{disclination density}. 

The \textit{dislocation density}, in the presence of disclinations is defined as
\[
\alpha_{ip} := - e_{pjk} u_{ijk} = - e_{pjk} \left(\p_k u_{ij} - e_{ijr} \omega^{(s)}_{rk} \right)
\]
and the curl of the elastic distortion satisfies the fundamental relation
\begin{equation*}
    e_{pkj} \p_k u_{ij} = \alpha_{ip} + \omega^{(s)}_{pi} - \omega^{(s)}_{kk} \delta_{ip} = \alpha^*_{ip}
\end{equation*}
which implies, after taking another curl and symmetrizing in the indices $r$ and $p$, the fundamental relation
\begin{equation}\label{eq:kroner_inc}
    e_{rqi}e_{pkj} \p_q \p_k \veps_{ij} = \frac{1}{2} \left[(e_{rqi} \p_q \alpha_{ip} + e_{pqi} \p_q \alpha_{ir} ) + (\theta_{rp} + \theta_{pr}) \right] = : s_{rp} \Longleftrightarrow \mbox{inc} \,\veps =  \overline{\mbox{curl}\, \left( \alpha^T \right)}  + \overline{\theta} =: s,
\end{equation}
where we use the notation
\begin{equation}\label{eq:not_(a)sym}
\begin{aligned}
\overline{(\cdot)}_{ij}   = \frac{1}{2} \left( (\cdot)_{ij} + (\cdot)_{ji} \right); & \qquad
\widetilde{(\cdot)}_{ij}  = \frac{1}{2} \left( (\cdot)_{ij} - (\cdot)_{ji} \right);\\ 
\overline{u}_{ij} =: \veps_{ij} & \qquad \widetilde{u}_{ij} =: r_{ij}.
\end{aligned}
\end{equation}
Since $\alpha^*$ is locally a curl, concentrations of this field along lines carry a topological charge and the (spatial part of the) current corresponding to the conservation of this charge is characterized by
\[
J_{ij} := e_{jrs} \alpha^*_{ir} V_s
\]
where $V_s$ is the velocity field convecting the defect lines of $\alpha^*$.
With this definition, \eqref{eq:strn_rate_decomp} can be written as 
\[
\p_j v_i - v_{ij} = v^{(s)}_{ij} := J_{ij}.
\]
\end{appendix}
\bibliographystyle{alpha}\bibliography{disloc_action_ref}
\end{document}